\documentclass[aps,prb,twocolumn,showpacs,superscriptaddress]{revtex4}

\usepackage{amsmath}
\usepackage{amsfonts}
\usepackage{amssymb}
\usepackage{graphicx}
\usepackage{dsfont}
\usepackage[justification=raggedright,singlelinecheck=false]{caption}

\newcommand{\abs}[1]{\ensuremath{\left\vert#1\right\vert}}

\newcommand{\arcsinh}[1]{\ensuremath{\operatorname{arcsinh}\left(#1\right)}}
\renewcommand{\arctan}[1]{\ensuremath{\operatorname{arctan}\left(#1\right)}}

\newcommand{\sign}[1]{\ensuremath{\operatorname{sign}\left(#1\right)}}
\newcommand{\heaviside}[1]{\ensuremath{\theta\left[#1\right]}}
\renewcommand{\Re}[1]{\ensuremath{\operatorname{Re} \left\{#1\right\} } }
\renewcommand{\Im}[1]{\ensuremath{\operatorname{Im} \left\{#1\right\} } }

\renewcommand{\vec}[1]{\boldsymbol{#1} }

\begin{document}

\title{Dielectric function, screening, and plasmons of graphene 
in the presence of spin-orbit interactions}
\author{Andreas Scholz}
\email[To whom correspondence should be addressed. Electronic
address: ]{andreas.scholz@physik.uni-regensburg.de}
\affiliation{Institute for Theoretical Physics, 
University of Regensburg, D-93040 Regensburg, Germany}
\author{Tobias Stauber}
\affiliation{Departamento de Fisica de la Materia Condensada 
and Instituto Nicolas Cabrera,
Universidad Autonoma de Madrid, E-28049 Madrid, Spain}
\author{John Schliemann}
\affiliation{Institute for Theoretical Physics, 
University of Regensburg, D-93040 Regensburg, Germany}
\date{\today}

\begin{abstract}
We study the dielectric properties of graphene in the presence of Rashba and 
intrinsic spin-orbit interactions in their most general form, i.e., for 
arbitrary frequency, wave vector, doping, and spin-orbit coupling (SOC) 
parameters. 
The main result consists in the derivation of closed analytical expressions for the imaginary as well as for the real part of the polarization function.
Several limiting cases, e.g., the case of purely Rashba or purely intrinsic 
SOC, and the case of equally large Rashba and intrinsic coupling parameters 
are discussed. 
In the static limit the asymptotic behavior of the screened potential due to 
charged impurities is derived. In the opposite limit ($q=0$, $\omega\to0$), an analytical expression for the plasmon dispersion is obtained and afterwards compared to the numerical result.
Our result can also be applied to related systems such as bilayer 
graphene or topological insulators.
\end{abstract}


\maketitle

\section{Introduction}
\label{sect:Introduction}
It is now well established that at low energies the charge carriers in 
graphene are described by a Dirac-like equation for massless 
particles.\cite{Wallace, CastroNetoRMP09}
While standard graphene, i.e., without any spin-orbit interactions (SOIs), does not exhibit a band 
gap, a gap opens up in the spectrum if one includes purely intrinsic 
spin-orbit interactions.\cite{Kane_2005}
The corresponding energy dispersion resembles that of a massive relativistic 
particle with a rest energy which is proportional to the spin-orbit 
coupling parameter (SOC).
Including SOIs of the Rashba type, e.g., by applying an external electric 
field, lifts the spin degeneracy.
Depending on the ratio of the intrinsic and the Rashba parameters a gap can 
occur in the spectrum or not.

Many theoretical studies on the dielectric function of various systems have 
been made in the last years.
Besides semiconductor two-dimensional electron gases\cite{Pletyuhkov_2006, Badalyan_2010, Ullrich_2003} 
and hole 
gas systems,\cite{Schliemann_2010} large investigations have been made in 
graphene.
Starting from the simplest possible graphene model within the Dirac-cone 
approximation,\cite{Wunsch, Sarma, Polini}
more and more extensions have been included. 
These extensions range from numerical\cite{Stauber_2010, Stauber_2010_2} and analytical\cite{Stauber_2010_3} tight-binding 
studies and the inclusion of a finite 
band gap \cite{Pyat, Wang_2007_2, Scholz_2011} to double- and multilayer graphene 
samples,\cite{Borghi_2009, Gamayun_2011, Yuan_2011, Stauber_2012,Profumo_2012, Triola_2012} graphene antidot
lattices,\cite{Schultz_2011} and graphene under a circularly polarized ac electric field.\cite{Busl_2011}

In this work we study the dielectric properties of graphene including both 
the Rashba and the intrinsic spin-orbit coupling.
While the case of purely intrinsic interactions is well 
understood,\cite{Pyat, Wang_2007_2, Scholz_2011} the dielectric function for the general case, where
both types of SOIs are present, is unknown.
Other previous studies have investigated the effect of SOI on 
magnetotransport \cite{Ding_2008} and the 
optical conductivity.\cite{Ingenhoven_2010,Hill_2011}

Our study is motivated by recent experimental and theoretical works 
demonstrating that the SOC parameters can significantly be enlarged by 
choosing proper adatoms \cite{Neto09,Weeks_2011, Ma_2012} or a suitable 
environment.\cite{Dedkov_2008, Varykhalov_2008, Jin_2012}

Information that can be extracted from the dielectric function range from the
screening between charged particles to the collective
charge excitations formed due to the long-ranged Coulomb interaction.
Knowledge of the latter is not only important for possible
future applications in the field of plasmonics, where graphene seems to be a promising
material,\cite{Plasmonics} but also because of fundamental reasons.
Recent experiments and theoretical studies showed that interactions between
charge carriers and plasmons in graphene, forming so-called plasmarons, yield to
measurable changes in the energy spectrum.\cite{Bostwock_2010}

\begin{figure*}
\includegraphics[scale=0.60]{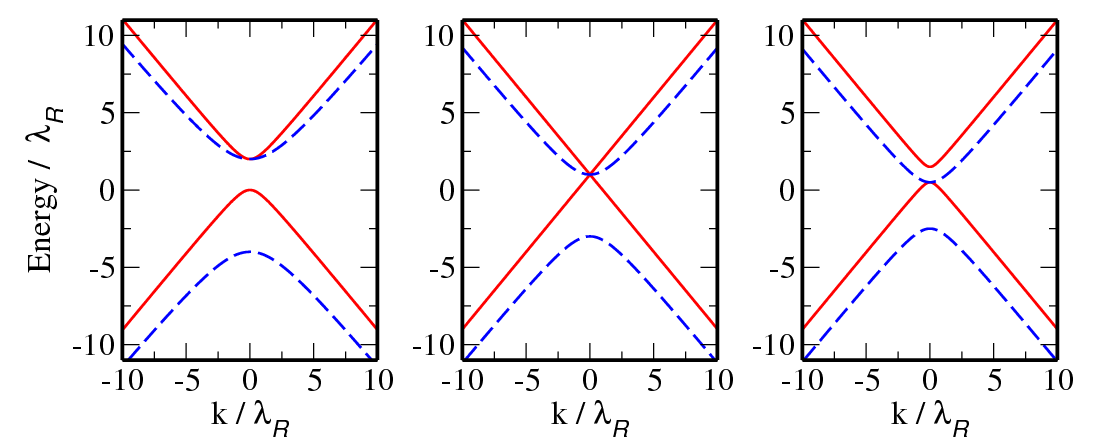}
\caption{Energy dispersion in units of $\lambda_R$ for $E_{+\pm}$ (solid lines) and $E_{-\pm}$ (dashed lines): (a) $\lambda_I=2\lambda_R$, (b) $\lambda_I=\lambda_R$, (c) $\lambda_I=\lambda_R/2$.}
\label{FIG_energy_dispersion}
\end{figure*}

The paper is organized as follows. In Sec. \ref{sect:Model}, we introduce 
the model Hamiltonian including the eigensystem and summarize the formalism
of the random phase approximation (RPA).
In Sec. \ref{sect:Results}, analytical and numerical results for the free 
polarization function of the undoped and the doped system are given.
In Sec. \ref{sect:Static}, the dielectric function is used to 
analyze the static screening properties due to charged impurities. 
We provide qualitatively the asymptotic behavior of the induced potential.
The long-wavelength collective charge excitations of graphene are derived 
in Sec. \ref{sect:Plasmons} and afterwards compared to the numerical result. 
We find the existence of several new potential plasmon modes that are absent without 
any spin-orbit interactions. Most of these zeros, however, are overdamped as can be seen
from the energy loss function.
We close with conclusions and outlook in Sec. \ref{sect:Conclusions}.
Finally, in Appendixes \ref{sect:AppA} and \ref{sect:AppB} we give details of 
the calculation of the free polarization function.

\section{The model}
\label{sect:Model}
We describe graphene with SOI within the Dirac cone approximation. At one $K$ point, the Hamiltonian is given by \cite{Kane_2005}
\begin{align}
\hat H = v_F \vec p \cdot \vec \tau + \lambda_R \left( \vec \tau \times \vec \sigma \right) \vec e_z + \lambda_I \tau_z \sigma_z .
\label{Hamiltonian}
\end{align}
The Pauli matrices $\vec\tau$ ($\vec\sigma$) act on the pseudospin (real spin) space.
The other $K$ point can be described by the above Hamiltonian with $\sigma_x\rightarrow-\sigma_x$ and $\sigma_z\rightarrow-\sigma_z$.
Since the two $K$ points are not coupled, we can limit our discussion to the above Hamiltonian, multiplying the final results by the valley index $g_v=2$. Moreover, without loss of generality, we assume a positive Rashba and intrinsic coupling as the eigensystem and thus the dielectric function is not changed for negative values.

\subsection{Solution}
For a sufficiently large intrinsic coupling parameter, $\lambda_I > \lambda_R$, the system is in the spin quantum Hall phase with a characteristic band gap.
For $\lambda_R > \lambda_I$ the gap in the spectrum is closed and the system behaves as an ordinary semimetal.
At the point where $\lambda_R = \lambda_I$ a quantum phase transitions occurs in the system. In the following we mainly set $v_F = 1$ and $\hbar = 1$.

The eigensystem reads
\begin{align}
&\left\vert \chi_{\pm\pm}(\vec k)\right\rangle = \frac 1{\sqrt2} \begin{pmatrix} \sin{\left(\theta_{\mp}/2\right)} \\  \cos{\left(\theta_{\mp}/2\right)} e^{i\varphi} \\ \pm \cos{\left(\theta_{\mp}/2\right)} e^{i\varphi} \\ \pm \sin{\left(\theta_{\mp}/2\right)} e^{2i\varphi} \end{pmatrix} , \notag \\ 
&\left\vert \chi_{\pm\mp}(\vec k)\right\rangle = \frac 1{\sqrt2} \begin{pmatrix} \cos{\left(\theta_{\mp}/2\right)} \\ - \sin{\left(\theta_{\mp}/2\right)} e^{i\varphi} \\ \mp \sin{\left(\theta_{\mp}/2\right)} e^{i\varphi} \\ \pm \cos{\left(\theta_{\mp}/2\right)} e^{2i\varphi} \end{pmatrix} ,
\end{align}
with $\sin{(\theta_\pm)} = \frac k{\sqrt{k^2+\lambda_{\pm}^2}}$ and $\lambda_{\alpha} = \lambda_R+\alpha\lambda_{I}$, and ($ k = \abs{\vec k}$)
\begin{align}
E_{\alpha\beta} (\vec k) = \alpha \; \lambda_R + \beta \; \sqrt{k^2 + {\lambda_{-\alpha}}^2 } \quad (\alpha,\beta = \pm1) . \label{Energy_dispersion}
\end{align}
For $\lambda_R \neq 0$ the spin degeneracy is lifted and two distinct Fermi wave vectors, 
$k_{F\pm} = \sqrt{\mu\left(\mu\mp2\lambda_R\right)\pm 2\lambda_R\lambda_I - \lambda_I^2}$, exist. In Fig. \ref{FIG_energy_dispersion}, the energy dispersion is shown for three characteristic values of the SOI.

The energy scales for the SOC parameters in monolayer graphene, $\lambda_I = 12 \mu {\rm eV}$ and $\lambda_R = 5 \mu {\rm eV}$ for an electric field of $1 {\rm\frac V{nm}}$, are generally small \cite{Gmitra_2009}.
However, it was shown that these parameters can be enlarged to $\lambda_I \approx 30 {\rm meV}$ for thallium adatoms \cite{Weeks_2011} or $\lambda_R \approx 13 \rm{meV}$ for graphene placed on a Ni(111) surface.\cite{Varykhalov_2008}

The above Hamiltonian with only Rashba coupling can be mapped onto the bilayer Hamiltonian without SOI, relating the interlayer hopping parameter $t_{IL} \approx 0.2 {\rm eV}$ \cite{Min_2007} to the Rashba SOC. Our findings can also be applied to a topological insulator within the Kane-Mele model.\cite{Kane_2005}

\begin{figure*}
\includegraphics[scale=0.50]{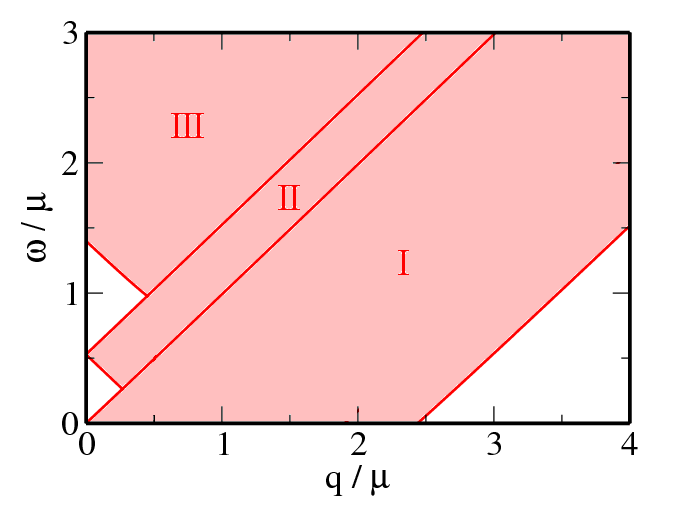}
\caption{Single-particle continuum (dark area) for the particular choice of $\lambda_R = 2\lambda_I = 0.3\mu$.
Analytical expressions for the boundaries of the distinct regions I, II and III can be found in Sec. \ref{sect:Results} B.
}
\label{FIG_SPE}
\end{figure*}

\subsection{Dielectric function}
In order to find the dielectric function in RPA \cite{Guil} given by
\begin{align}
\varepsilon(q,\omega) = 1 - V(q) \chi_0(q,\omega), 
\end{align}
where $V(q) = \frac{e^2}{2\epsilon_0 q}$ is the Fourier transform of the Coulomb potential in two dimensions,
$V(r) = \frac{e^2}{4\pi\epsilon_0 r}$, and $\epsilon_0$ the vacuum permittivity,
one needs to calculate the free polarization,
\begin{align}
&\chi_0(q,\omega) = \sum_{\alpha,\eta_i=\pm1} \int \frac {g_vd^2k}{(2\pi)^2} \left|\left\langle\chi_{\eta_1\eta_2}\left(\vec k\right) \bigg| \chi_{\eta_3\eta_4}\left(\vec k+\vec q\right) \right\rangle\right|^2 \notag \\
& \times \frac {\alpha \, f(E_{\eta_1\eta_2}(\vec k))} {\omega - \alpha \left[E_{\eta_3\eta_4}\left(\vec k+\vec q\right) - E_{\eta_1\eta_2}\left(\vec k\right) \right] + i0}. \label{DEF_suscept}
\end{align}
In the following we assume zero temperature. The Fermi function $f(E)$ then reduces to a simple step function.
Because of the general relation $\chi_0(q,-\omega) = \left[ \chi_0(q,\omega) \right]^*$, we restrict our discussions to positive frequencies $\omega$. 

\begin{figure*}
\includegraphics[scale=0.55]{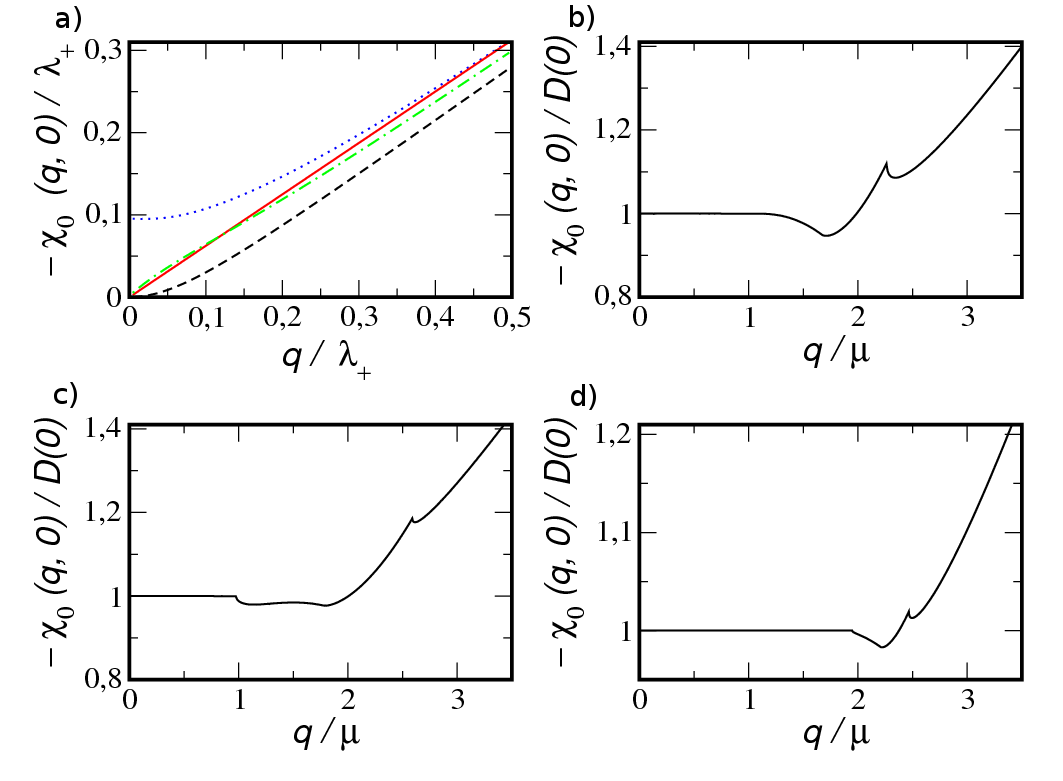}
\caption{Static real part of the charge susceptibility for
(a) undoped graphene with fixed $\lambda_+$ for $\lambda_R = 2\lambda_I$ (dotted),
$\lambda_R = \lambda_I$ (dot-dashed), $\lambda_I = 2\lambda_R$ (dashed), $\lambda_R = \lambda_I = 0$ (straight),
and doped graphene with (b) $\lambda_R = \lambda_I = 0.3 \mu$,
(c) $\lambda_R = 2 \lambda_I = 0.3 \mu$, and
(d) $2 \lambda_R = \lambda_I = 0.3 \mu$ in units of the density of states $\rm{D}(0) = g_v\mu/\pi$.
}
\label{FIG_Im_part_int}
\end{figure*}

\section{Analytical results}
\label{sect:Results}
\subsection{Zero doping}
For zero doping the valence bands are completely filled while the conduction bands are empty. Only transitions between bands
$E_{\alpha -}$ and $E_{\beta +}$ are possible. The resulting charge correlation function can be decomposed as
\begin{align}
\bar\chi_0(q,\omega) = \sum_{\eta_i=\pm} \chi^{\eta_1-\to\eta_3+}(q,\omega)
\end{align}
Here we introduced the notation $\chi^{\eta_1\eta_2\to\eta_3\eta_4}(q,\omega)$ describing transitions
from the initial band $E_{\eta_1\eta_2}(\vec k)$ to the final band $E_{\eta_3\eta_4}(\vec k + \vec q)$.
For the imaginary part we find
\begin{align}
&\Im{\chi^{\mp-\to\mp+}(q,\omega)} = \frac {g_v}{16} \; \heaviside{\omega^2 - q^2 - 4{\lambda_\pm}^2} \notag \displaybreak[0]\\
& \quad \times \left[ \frac{3q^4-4{\lambda_\pm}^2q^2 - 5q^2\omega^2 + 2\omega^4}{\left(\omega^2-q^2\right)^{3/2}} \right. \notag \\
& \quad \left. - \frac{\abs{q^2-\omega\left(\omega-2\lambda_\pm\right)} + \abs{q^2-\omega\left(\omega+2\lambda_\pm\right)}}{\omega} \right]
\label{int_Im_part_2}
\end{align}
and
\begin{align}
&\Im{\chi^{\pm-\to\mp+}(q,\omega)} = -\frac {g_v}8 \, \heaviside{\omega_\pm^2-q^2-4\gamma^2} \times \\
& \left[ \sqrt{\omega_\pm^2-q^2} 
- \frac{\abs{q^2-\omega\left(\omega\pm2\lambda_-\right)}}{2\omega} 
- \frac{\abs{q^2-\omega\left(\omega\pm2\lambda_+\right)}}{2\omega}
\, \right] \notag
\end{align}
Here we defined  $\omega_\pm = \omega \pm 2\lambda_R$ and $\gamma = \text{max}\{\lambda_R,\lambda_I\}$.
For equally large spin-orbit coupling parameters, $\lambda_R=\lambda_I$, the imaginary part is divergent at the threshold $\omega = q$ but finite otherwise. The divergent part of the polarization is $\chi^{+-\to++}$ as the bands $E_{+\pm}(\vec k)$ are linear in momentum.

The real part can be obtained via the Kramers-Kronig relations
\begin{align}
\Re{\bar\chi_0(q,\omega)} = \frac 2\pi \mathcal P \int_0^\infty d\omega' \frac {\omega' \Im{\bar\chi_0(q,\omega')}}{\omega'^2-\omega^2} . \label{KKR}
\end{align}
After carrying out the remaining integration, where it is necessary to keep the principal value, we arrive at
\begin{widetext}
\begin{align}
&\Re{\chi^{\mp-\to\mp+}(q,\omega)} = \frac {g_v}{8\pi} \left[ - 2\lambda_\pm + 2\sqrt{q^2+\lambda_\pm^2} 
 \left(5q^2\omega^2+4q^2\lambda_\pm^2 - 3q^4 - 2w^4\right) \Re{\frac{\arctan{\frac{\sqrt{q^2-\omega^2}}{2\lambda_\pm}}}{\left(q^2-\omega^2\right)^{3/2}} } \right. \notag \\
& \left. - \frac{2q^2\lambda_\pm}{q^2-\omega^2} 
+ 2\lambda_\pm \ln{\abs{\frac{q^2-\omega^2+4\lambda_\pm^2}{\left(\sqrt{q^2+\lambda_\pm^2}+\lambda_\pm\right)^2-\omega^2}}} 
- \frac{\omega^2-q^2}{2\omega} \ln{\frac{\sqrt{q^2+\lambda_\pm^2}+\lambda_\pm + \omega}{\sqrt{q^2+\lambda_\pm^2}+\lambda_\pm - \omega}}
\right]
\end{align}
\end{widetext}
and
\begin{figure*}
\includegraphics[scale=0.50]{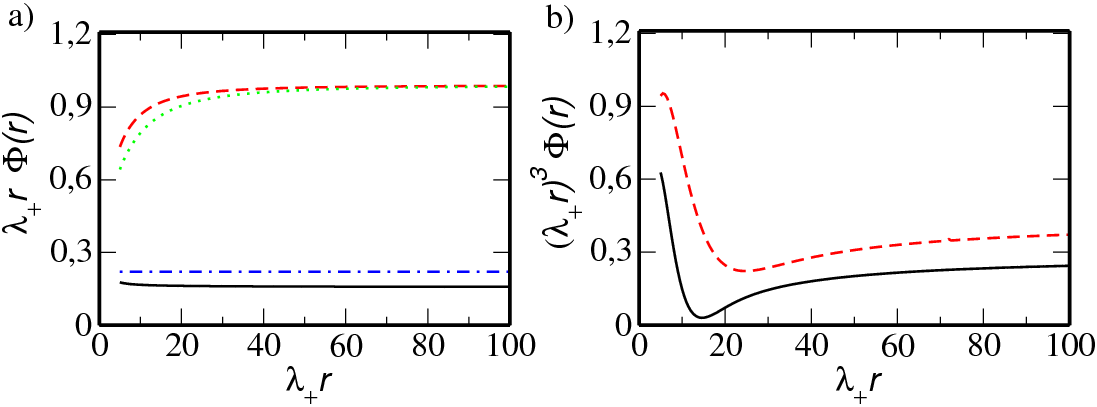}
\caption{Asymptotic screened potential (in units of $Q\lambda_+/\epsilon_0$) of undoped graphene
for fixed $\lambda_+=\lambda_R+\lambda_I$ and different spin-orbit coupling parameters.
(a) $\lambda_R = \lambda_I$ (straight line), $\lambda_I = 2\lambda_R$ (dashed), $\lambda_R=0$ (dotted). Also shown the non-interacting case $\lambda_R = \lambda_I=0$ (dot-dashed). (b) $\lambda_I=0$ (straight), $\lambda_R = 2\lambda_I$ (dashed).
}
\label{FIG_screened_potential_int}
\end{figure*}
\begin{widetext}
\begin{align}
&\Re{\chi^{\pm-\to\mp+}(q,\omega)} = -\frac {g}{4\pi} \left[ 
2\left(\pm\lambda_R - \gamma \pm\lambda_R \ln{4} \right) \mp 2\lambda_R \arcsinh{\frac{2\gamma}q} \right. \notag \displaybreak[0] \\
& \left. -\frac 12 \Re{\sqrt{q^2-\omega_\pm^2} \arcsin{\frac{\sqrt{q^2-\omega_\pm^2}\left[q+\omega_\pm\left(2\gamma - \sqrt{q^2+4\gamma^2}\right) \right] }{\sqrt{q^2+4\gamma^2}-\omega_\pm}} } \right. \notag \displaybreak[0] \\
& \left. -\frac 12 \Re{ 
\sqrt{q^2-\omega_\mp^2} \arcsin{\frac{\sqrt{q^2-\omega_\mp^2}\left[q-\omega_\mp\left(2\gamma - \sqrt{q^2+4\gamma^2}\right) \right] }{\sqrt{q^2+4\gamma^2}+\omega_\mp}} } \right. \notag \displaybreak[0] \\
& \left. + \heaviside{\pm\lambda_\pm} \mathcal L_{(\pm\lambda_\mp)}\left(\sqrt{q^2+\lambda_\mp^2}\mp\lambda_\mp\right)
- \frac 12 \sign{\pm\lambda_\pm} \mathcal L_{(\pm\lambda_\mp)}\left(\sqrt{q^2+4\gamma^2}\mp2\lambda_R\right) \right. \notag \displaybreak[0] \\
& \left. + \heaviside{\pm\lambda_\mp} \mathcal L_{(\pm\lambda_\pm)}\left(\sqrt{q^2+\lambda_\pm^2}\mp\lambda_\pm\right)
- \frac 12 \sign{\pm\lambda_\mp} \mathcal L_{(\pm\lambda_\pm)}\left(\sqrt{q^2+4\gamma^2}\mp2\lambda_R\right)
\right] \displaybreak[0] 
\end{align}
\end{widetext}
Here we defined the function
\begin{align}
\mathcal L_{\lambda}(x) = x + \lambda \ln{\frac{x^2-\omega^2}{q^2}} - \frac{\omega^2-q^2}{2\omega}\ln{\abs{\frac{x+\omega}{x-\omega}}} .
\end{align}
\begin{figure*}
\includegraphics[scale=0.50]{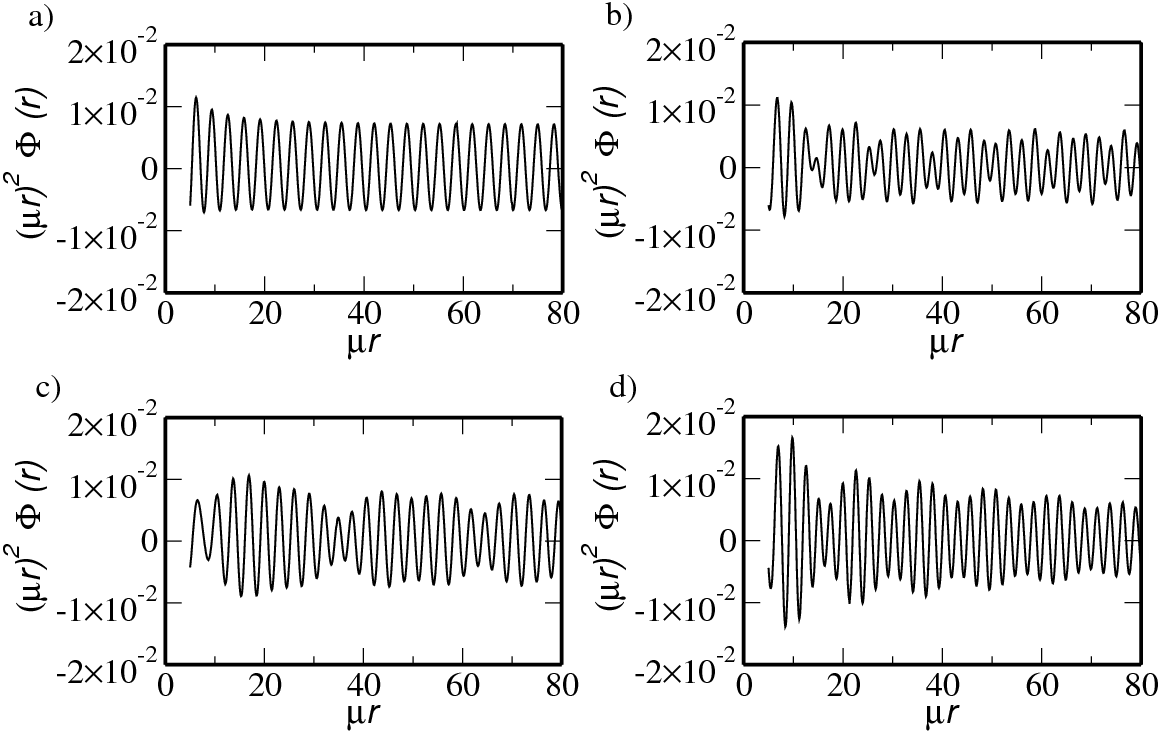}
\caption{Asymptotic screened potential (in units of $Q\mu/\epsilon_0$) of doped graphene for various spin-orbit coupling parameters:
(a) $(\lambda_R/\mu, \lambda_I/\mu) = (0,0.3)$, (b) $(0.3,0.15)$,
(c) $(0.15,0.3)$, and (d) $(0.3,0.3)$.
}
\label{FIG_ext_screening}
\end{figure*}

\subsection{Finite doping}
We now continue with the case of a finite chemical potential lying in the conduction band (the $p$-doped case is analogous). The free polarization in the doped case reads
 \begin{align}
\chi_0(q,\omega) = \bar\chi_0(q,\omega) + \delta\chi_{k_{F+}} (q,\omega) + \delta\chi_{k_{F-}} (q,\omega).
 \end{align}
$\bar\chi_0$ is the undoped part given above.
The two remaining contributions, $\delta\chi_{k_{F+}}$ and $\delta\chi_{k_{F-}}$, with
\begin{widetext}
\begin{align}
\delta\chi_{k_{F\pm}}(q,\omega) = \frac{g_v}{4\pi^2} \sum_{\alpha, \mu,\nu=\pm1} \mathcal P \int_0^{k_{F\pm}} d^2k \sum_{\alpha=\pm1}
\frac {\alpha \left\vert \left\langle \chi_{\pm+}(\vec k) \Big\vert \chi_{\mu\nu}(\vec k+\vec q) \right\rangle \right\vert ^2}
{\omega + i0 - \alpha \left[ E_{\mu\nu}(\vec k+\vec q) - E_{\pm+}(\vec k) \right] } 
\end{align}
\end{widetext}
refer to transitions with initial states in band $E_{++}$ and $E_{-+}$, respectively.
As the expressions for the extrinsic real and imaginary part of the free polarization function are quite lengthy,
we refer to Appendix \ref{sect:AppB} where the results including major steps of the derivation can be found.

Similar to the undoped case, the density correlation function of graphene is finite at $\omega = q$ for $\lambda_R \neq \lambda_I$
and divergent for $\lambda_R=\lambda_I$. However, this divergence vanishes in the RPA improved result.\cite{Wunsch}

From the shape of the Fermi surface and the dispersion relation as given in Eq. (\ref{Energy_dispersion}), we can determine
the boundaries of the dissipative electron-hole continuum.\cite{Guil}
In Fig. \ref{FIG_SPE} this is shown for the particular choice of $\lambda_R = 2\lambda_I = 0.3\mu$.
In general, the lower and upper boundaries of the damped region I are given by
\begin{align*}
\omega_{low}^I  = \text{max} \left\{ 0, \; \sqrt{\left(k_{F-}-q\right)^2 + \lambda_+^2} - \sqrt{k_{F-}^2 + \lambda_+^2} \right\}
\end{align*}
and 
\begin{align*}
\omega_{up}^I = \text{max} &\left\{ \sqrt{\left(k_{F-}+q\right)^2 + \lambda_+^2} - \sqrt{k_{F-}^2 + \lambda_+^2} , \right. \\
& \left. \sqrt{\left(k_{F+}+q\right)^2 + \lambda_-^2} - \sqrt{k_{F+}^2 + \lambda_-^2}
 \right\} ,
\end{align*}
respectively. Region I is due to intraband transitions from band $E_{\pm+}(\vec k)$ to $E_{\pm+}(\vec k+\vec q)$.
Region \textit{II} accounts for interband transitions between conduction bands and is confined by
\begin{align*}
\omega_{up/low}^{II} = \sqrt{\left(k_{F-}\pm q\right)^2 + \lambda_-^2} - \sqrt{k_{F-}^2 + \lambda_+^2} + 2\lambda_R .
\end{align*}
For region III the lower limit reads
\begin{align*}
\omega_{low}^{III} = \text{min} & \left\{ \sqrt{\left(k_{F-}-q\right)^2 + \lambda_+^2 } + \sqrt{k_{F-}^2 + \lambda_-^2 } 
- 2 \lambda_R , \right. \\
& \left. \sqrt{\left(k_{F+}-q\right)^2 + \lambda_-^2 } + \sqrt{k_{F+}^2 + \lambda_-^2 }
\right\}
\end{align*}
while there is no restriction to the upper boundary.
This part is due to transitions between valence and conduction bands.
\begin{figure*}
\includegraphics[scale=0.60]{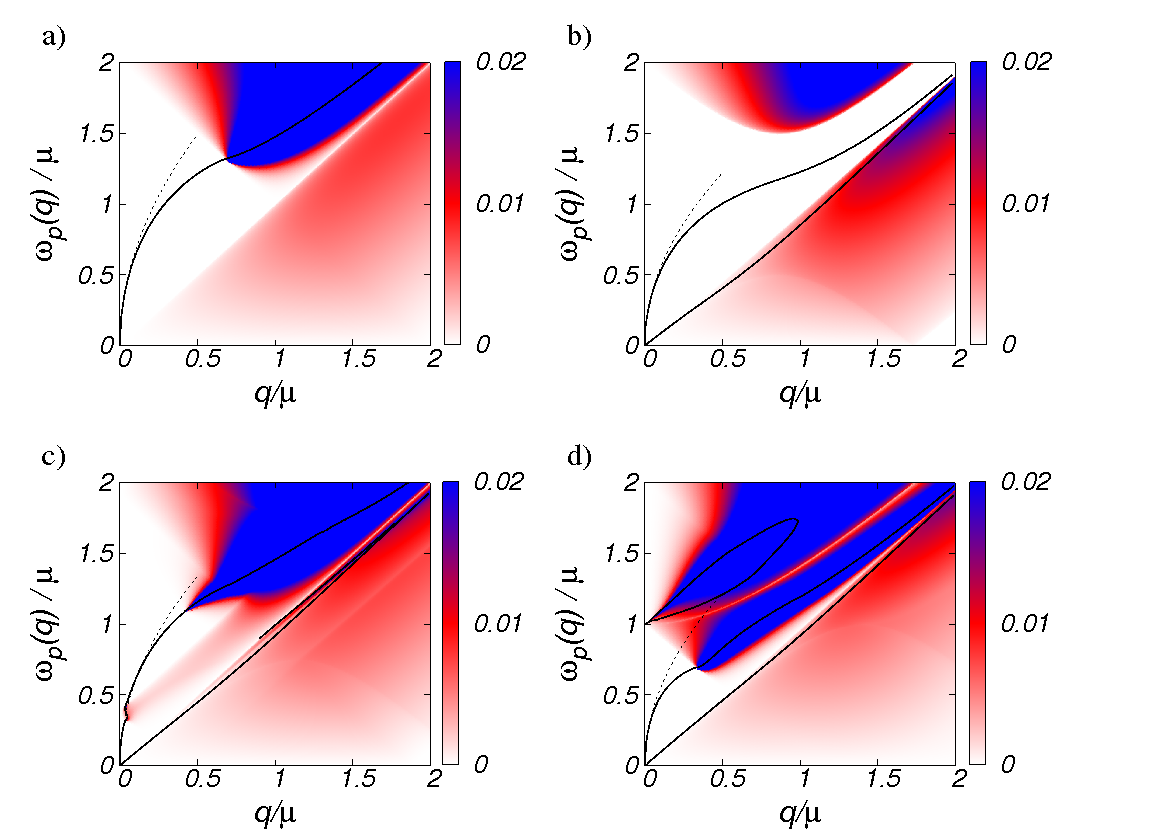}
\caption{Density plot of $\Im{-1/\varepsilon(q,\omega)}$ for various spin-obit coupling parameters: (a) $(\lambda_R/\mu,\lambda_I/\mu) = (0,0)$, (b) $(0,0.5)$, (c) $(0.25, 0.25)$, (d) $(0.5, 0)$. Straight lines correspond to the numerically calculated zeros of the real part of the dielectric function while dashed lines represent the long-wavelength result of Eq. (\ref{small_q_plasmons}).}
\label{FIG_plasmon_dispersion}
\end{figure*}

\section{Screening of impurities}
\label{sect:Static}
The potential of a screened charged impurity is obtained from the definition of the dielectric function,
\begin{align}
\Phi(r) = \frac Q{\epsilon_0} \int_0^\infty dq \; \frac{J_0(qr)}{\varepsilon(q,0)} . \label{screened_pot}
\end{align}
$J_0(x)$ is the Bessel function of the first kind and $Q$ the charge of the impurity.
Making use of Eq. (\ref{screened_pot}) the screened potential for the undoped system is calculated numerically where
$\Phi(r)$ is mainly determined by the long-wavelength behavior of the static correlator.\cite{Gamayun_2011}
As can be seen from Fig. \ref{FIG_Im_part_int}(a), the long-wavelength limit of the polarization $\bar\chi_0(0,0)$ is finite in the semimetallic state ($\lambda_R>\lambda_I$) and zero otherwise while for large momenta all functions scale like $1/q$.
From Fig. \ref{FIG_screened_potential_int}(a) we can see that for $\lambda_I \gtrsim \lambda_R$ the potential scales like $\Phi(r) \propto 1/r$ at large distances.
For $\lambda_R > \lambda_I$ the asymptotic potential behaves as $\Phi(r) \propto 1/r^3$; see Fig. \ref{FIG_screened_potential_int}(b).
The actual values of $\mu r$ at which the above asymptotics are appropriate approximations depend on the difference of $\lambda_R$ and $\lambda_I$.
As mentioned in the introduction, the two different parameter regimes belong to different phases separated by the quantum critical point at $\lambda_R = \lambda_I$.

\indent The static density correlator for the doped system is much more complicated.
Integrals of the form (\ref{screened_pot}) are usually treated analytically by approximating the Bessel function by its asymptotic values.
The subsequent Fourier integral can then be solved with the Lighthill theorem.\cite{Lighthill}
The above theorem states that singularities in the derivatives of the dielectric function give rise to a characteristic, algebraic, oscillating decay of the screened potential. Physically, these Friedel oscillations are due to backscattering on the Fermi surface. We can thus make qualitative predictions for the potential $\Phi(r)$ at large distances away from the impurity, only from the analytical structure of the polarization function without carrying out the integration. Afterwards these predictions are compared to the exact numerical solution.
\begin{figure*}
\includegraphics[scale=0.6]{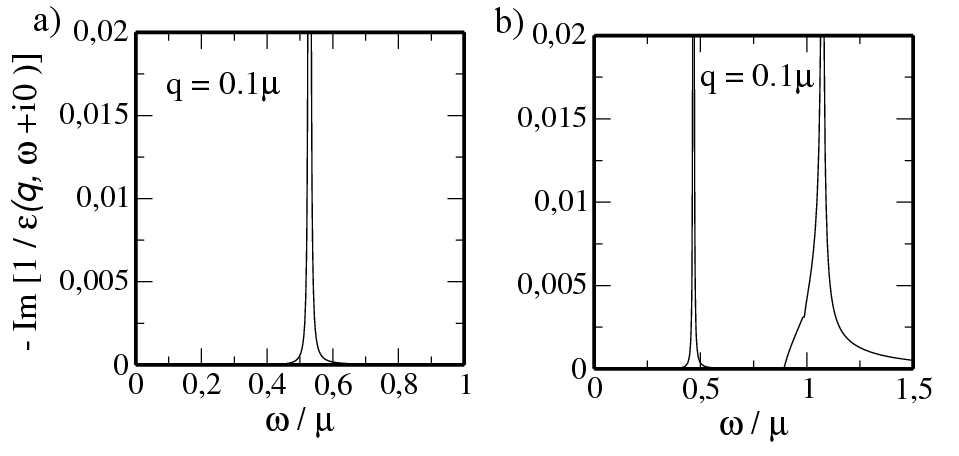}
\caption{Energy loss function $\Im{-1/\varepsilon(q,\omega+i0)}$ for fixed $q=0.1\mu$ with (a) $\lambda_R=0$, $\lambda_I=0.5\mu$ and (b) $\lambda_R=0.5\mu$, $\lambda_I=0$.}
\label{FIG_loss_function}
\end{figure*}

For non zero SOC and $\lambda_R\neq\lambda_I$ the first derivative of the polarization function is singular at special points $q=2k_{F\pm}$; see Figs. \ref{FIG_Im_part_int}(c) and (d). According to the Lighthill theorem the potential will exhibit a superposition of two different kinds of oscillations whereat $\Phi(r) \propto 1/r^2$.
This beating should be observable in sufficiently clean samples if the Rashba parameter, and the consequential breaking
of the spin-degeneracy, is large enough.
For predominant intrinsic SOI, the two oscillatory parts interfere constructively finally yielding an additional spin-degeneracy factor of $g_s=2$.\cite{Pyat}
For $\lambda_R = \lambda_I$ already the first derivative of $\chi_0(q,0)$ is singular at $q=2k_{F-}$ while at $q=2k_{F+}$ only the second derivative diverges; see Fig. \ref{FIG_Im_part_int}(b).
The main contribution in the potential again will be of order $1/r^2$.
The numerical inspection of $\Phi(r)$ as displayed in Fig. \ref{FIG_ext_screening} confirms the above predictions.

The resulting potential deviates significantly from the 
\begin{align}
\Phi(r) \propto \frac{\cos{\left(2k_{F}r\right)}}{\left(2k_{F}r\right)^3}
\end{align}
behavior of standard graphene within the Dirac cone approximation.\cite{Wunsch, Ando_2006}

Nevertheless, including the full dispersion of graphene can also lead to a different decay behavior, i.e., to anisotropic regular Friedel oscillations decaying like $1/r^2$.\cite{Santos_2011}

\section{Plasmons}
\label{sect:Plasmons}
Plasmons are defined as the zeros of the dielectric function,
\begin{align}
\varepsilon (q, \omega_p - i\gamma) = 0 . \label{Def_Plasmon_exact}
\end{align}
For small damping constant $\gamma$, Eq. (\ref{Def_Plasmon_exact}) can be substituted by the approximate equation\cite{Guil}
\begin{align}
\Re{\varepsilon (q, \omega_p)} = 0 . \label{Def_Plasmon_small_damping}
\end{align}
Only if $\gamma$ is small compared to $\omega_p$, one can speak of collective density fluctuations. For large Landau-damping, it is thus important to also discuss the more general energy loss function $\Im{-1/\varepsilon(q,\omega)}$ which gives the spectral density of the internal excitations of the system.

Similar to Refs. \cite{Pyat,Gamayun_2011}, there are several solutions of Eq. (\ref{Def_Plasmon_small_damping}) for non zero SOC parameters. In Fig. \ref{FIG_plasmon_dispersion}, these solutions are shown as straight lines together with a density plot of the energy loss function $\Im{-1/\varepsilon(q,\omega)}$. One of these solutions has an almost linear dispersion with a sound velocity close to the Fermi velocity which exhibits an ending point for $\lambda_R\sim\lambda_I$ associated with a double zero of the real part of the dielectric function. However, as can be seen from Figs. \ref{FIG_loss_function}(a) and (b), this solution does not yield to a resonance in the loss function and does thus not resemble a plasmonic mode. In the case where the gap in the spectrum is closed ($\lambda_R>\lambda_I$),
two additional zeros appear leading to potential high energy modes similar to bilayer graphene.\cite{Gamayun_2011, Yuan_2011}
However, these potential collective modes are damped by interband transitions;
i.e., the corresponding peaks in the loss function are broadened out as can be seen from Fig. \ref{FIG_loss_function}(b) and no clear signature is seen in the density plot.

We are thus left with the branch which is also present for ``clean'' graphene and which resembles the only genuine plasmonic mode; see Fig. \ref{FIG_plasmon_dispersion}(a). Its dispersion $\omega_p$ can be approximated in the long-wavelength limit ($q \ll \omega$) by\cite{Sensarma_2010}
\begin{align}
\omega_p^{0}(q) = \beta \sqrt{q},
\label{small_q_plasmons}
\end{align}
where the prefactor is given by $\beta = \sqrt{\frac{g_ve^2}{8\pi\epsilon_0} \sum_{\mu=\pm1} \frac{k_{F\mu}^2}{\sqrt{k_{F\mu}^2 + \lambda_{-\mu}^2}} }$.
We thus recover the typical $\sqrt{q}$-dispersion of 2D-plasmons.
\begin{figure*}
\includegraphics[scale=0.55]{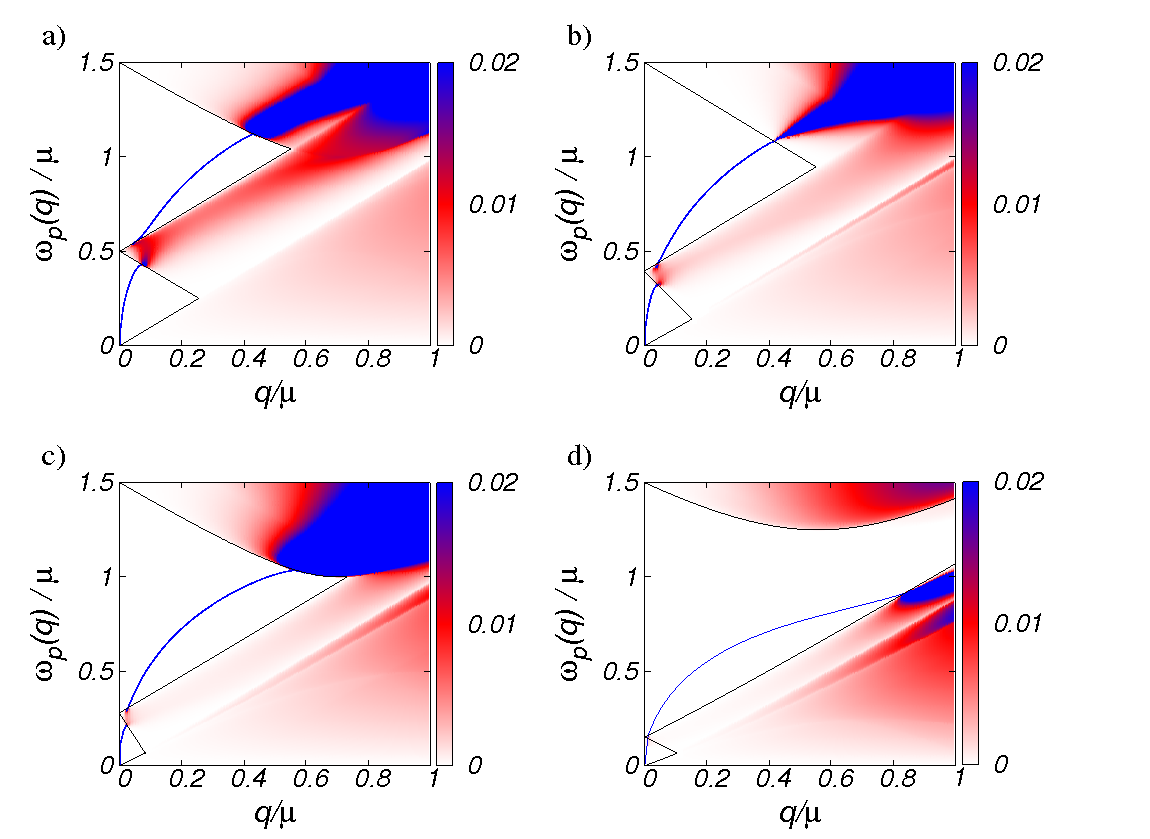}
\caption{Energy loss function $\Im{-1/\varepsilon(q,\omega+i0)}$ for (a) $(\lambda_R/\mu,\lambda_I/\mu)=(0.25,0)$, (b) $(0.25,0.25)$, (c) $(0.25,0.5)$, and (d) $(0.25,0.75)$. The straight blue lines show the undamped plasmon modes.
The black lines indicate the boundaries of the single-particle continuum (see Sec. III B).
}
\label{FIG_spectral_weight}
\end{figure*}

The long-wavelength approximation is shown as a dashed line in Fig. \ref{FIG_plasmon_dispersion} and coincides with the numerical solution, $\omega_p$, for small momenta, whereas for larger momenta, the $\omega_p$ is red shifted compared to $\omega_p^0$. If $\lambda_I$ is large enough, $\omega_p$ remains in the region where Landau damping is absent, see Fig. \ref{FIG_plasmon_dispersion}(b),\cite{Pyat} otherwise it eventually enters the Landau-damped region due to interband transitions from the valence to the conduction band, see Figs. \ref{FIG_plasmon_dispersion}(a), (d).

For two occupied conduction bands, which is the case in Fig. \ref{FIG_plasmon_dispersion}(c) and in Fig. \ref{FIG_spectral_weight}, the plasmon mode is disrupted at $q\approx 0.05\mu$ by a region with a finite imaginary part where it becomes damped. This additional Landau-damped region is due to interband transitions from the two conduction bands. The analytical description of the boundaries of this region can be found in Sec. \ref{sect:Results} B.

\begin{figure*}
\includegraphics[scale=0.45]{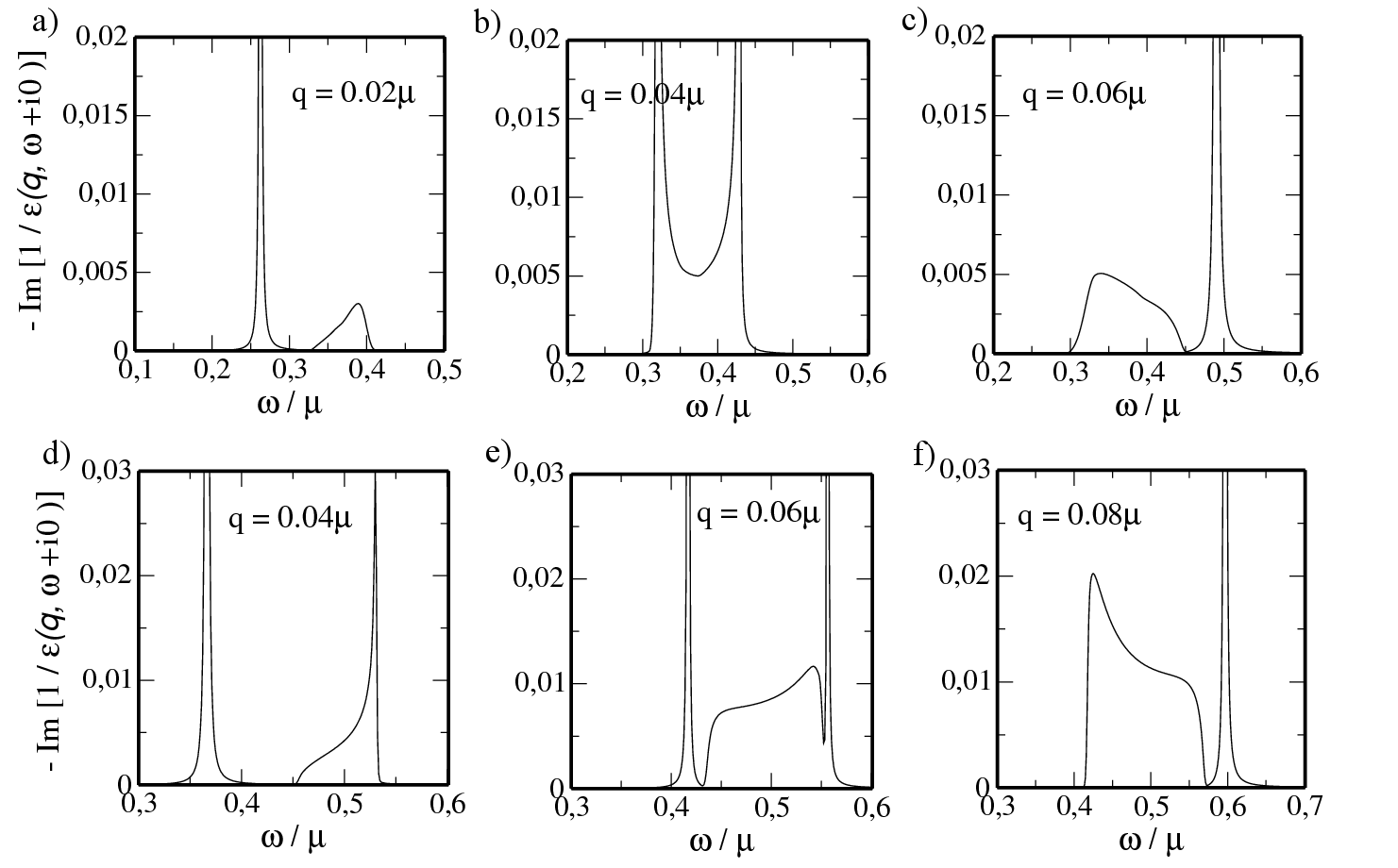}
\caption{Top: Energy loss function $\Im{-1/\varepsilon(q,\omega+i0)}$ for $\lambda_R = 0.25\mu = \lambda_I=0.25\mu$ 
and various wave vectors q. Bottom: The same for $\lambda_R=0.25\mu$ and $\lambda_I=0$.}
\label{FIG_loss_function2}
\end{figure*}

This ``pseudo gap'' of the plasmon dispersion can also be obtained from only considering Eq. ($\ref{Def_Plasmon_small_damping}$) since the ``plasmon'' velocity formally diverges at the entering and exit point as can be seen from Fig. \ref{FIG_plasmon_dispersion}(c).
The crossing points can alternatively be approximated by looking at the intersection of this region with the analytical long-wavelength  
approximation of the analytic plasmon dispersion. 
For the quantum critical point ($\lambda_R = \lambda_I$), this leads to the critical wave vector
\begin{align*}
q_{cr}^{\pm} = \frac {\left(\beta - \sqrt{\beta^2 \mp 4\left(k_{F-} - \sqrt{k_{F-}^2 + 4\lambda_R^2} + 2\lambda_R\right)} \right)^2} 4 \, ,
\end{align*}
and in particular to $q^-_{cr} \approx 0.019\mu$ and $q^+_{cr} \approx 0.025\mu$ for $\lambda_{R/I} = 0.25\mu$.
For a proper analysis, the full energy loss function thus needs to be discussed which is done in Fig. \ref{FIG_loss_function2}. It shows how the spectral weight is eventually transferred from the lower to the upper band as momentum is increased, explaining the step in the plasmon spectrum as shown in Fig. \ref{FIG_spectral_weight}. 

The pseudo gap of the plasmonic mode always appears for $\lambda_R < 0.5\mu$, since then two conduction bands are occupied independently of the value of $\lambda_I$, but it decreases for increasing $\lambda_I$ as the dissipative region due to interband transitions diminishes. In the opposite case of $\lambda_R > 0.5\mu$, either one or two bands can be occupied. For zero intrinsic coupling, the pseudo-gap is absent but increases up to a maximum value at around $\lambda_I \approx \lambda_R$ for increasing $\lambda_I$.

\indent Let us close with a comment on plasmons in undoped graphene. For neutral monolayer graphene and at zero temperature, plasmons can exist if one takes into account a circularly polarized light field\cite{Busl_2011} or effects beyond RPA,\cite{Mishchenko_2008} and in bilayer by including trigonal warping.\cite{Wang_2007} In our system, the real part of the dielectric function is always nonzero for the undoped case and thus no plasmons exist.

\section{Conclusions and outlook}
\label{sect:Conclusions}
We have presented analytical and numerical results for the dielectric function of monolayer graphene in the presence of Rashba and intrinsic spin-orbit interactions within the random phase approximation for finite frequency, wave vector, and doping.
The cases of predominant Rashba and intrinsic coupling and the case of equally large SOC were opposed.

In the static limit the screening properties due to external impurities were studied.
Our findings show that the power-law dependence of the screened potential in the undoped system depends on the ratio of the Rashba and intrinsic parameters.
While for $\lambda_R > \lambda_I$ the screened potential scales like $\Phi(r) \propto 1/r^3$,
for $\lambda_I \ge \lambda_R$ a weaker screening, $\Phi(r) \propto 1/r$, was found.
For finite Rashba coupling, a beating of Friedel oscillations in the doped system occurs due to
the existence of two distinct kinds of Fermi wave vectors. For large $\lambda_I \gg \lambda_R$, this beating vanishes 
and the two contributions interfere constructively.

In the last section the influence of SOI on the collective charge excitations was discussed. We found that while only one plasmon mode exists for standard graphene, several new potential modes occur for finite SOC.
However, most of these modes are overdamped and can hardly be detected as they lie in the region with finite Landau damping.
In the case when the two conduction bands are filled, the undamped plasmon mode is disrupted by a narrow dissipative region strip due to particle-hole excitations. This ``pseudo gap'' might be useful to gain further control in possible plasmonic circuitries based on graphene.

As already mentioned in the beginning, our findings go even beyond monolayer graphene.
For purely Rashba coupling the dielectric function presented in this work equals that of bilayer graphene.
The role of the SOC parameter is then played by the interlayer hopping amplitude $t_{IL}$ being several orders of magnitude larger than $\lambda_R$.
Additionally, the Hamiltonian in Eq. (\ref{Hamiltonian}) generally describes a system known as the Kane-Mele topological insulator.\cite{Kane_2005} Our discussion can thus be fully adopted to materials modeled by this Hamiltonian.
Besides that, our findings might also be relevant for other monolayers with similar symmetry properties compared to those of graphene,
e.g. MoS$_2$, where SOC is naturally strong.\cite{Xiao_2012} A detailed study of the dielectric properties of MoS$_2$, however, is left open for future works.

\acknowledgments
We thank G. G\'omez-Santos and O.~V. Gamayun for useful discussions and comments. This work was supported by Deutsche Forschungsgemeinschaft via Grant No. GRK 1570, by FCT under Grant No. PTDC/FIS/101434/2008, and MIC under Grant No. FIS2010-21883-C02-02.

\appendix

\begin{widetext}
\section{Details of the calculation of the undoped polarization}
\label{sect:AppA}
The undoped polarization is composed of four parts,
\begin{align}
\bar\chi_0(q,\omega) = \sum_{\eta_i=\pm} \chi^{\eta_1-\to\eta_3+}(q,\omega) .
\end{align}
As two of them can be obtained by a simple substitution, i.e.,
$\chi^{--\to++}_{\lambda_R} (q,\omega) = \chi^{+-\to-+}_{-\lambda_R} (q,\omega)$ and $\chi^{+-\to++}_{\lambda_R} (q,\omega) = \chi^{--\to-+}_{-\lambda_R} (q,\omega)$
only two contributions remain. With the help of the Dirac identity the imaginary parts read
\begin{align}
&\Im{\chi^{--\to-+} (q,\omega)} = -\frac{g_v}{4\pi} \int d^2k \sum_{\alpha=\pm1} \alpha
\left\vert \left\langle \chi_{--}(\vec k) \Big\vert \chi_{-+}(\vec k+\vec q) \right\rangle \right\vert ^2
\delta\left[\omega - \alpha \left( E_{-+}(\vec k+\vec q) - E_{--}(\vec k) \right)\right] \\
&= \frac {g_v}{16} \; \heaviside{\omega^2 - q^2 - 4{\lambda_+}^2} 
\left[ \frac{3q^4-4{\lambda_+}^2q^2 - 5q^2\omega^2 + 2\omega^4}{\left(\omega^2-q^2\right)^{3/2}} 
- \frac{\abs{q^2-\omega\left(\omega-2\lambda_+\right)} + \abs{q^2-\omega\left(\omega+2\lambda_+\right)}}{\omega} \right]
\end{align}
and
\begin{align}
&\Im{\chi^{+-\to-+} (q,\omega)} = -\frac{g_v}{4\pi} \int d^2k \sum_{\alpha=\pm1}
\alpha \left\vert \left\langle \chi_{+-}(\vec k) \Big\vert \chi_{-+}(\vec k+\vec q) \right\rangle \right\vert ^2
\delta\left[ \omega - \alpha \left( E_{-+}(\vec k+\vec q) - E_{+-}(\vec k) \right) \right] \\
&= -\frac{g_v}{8\pi} \int_{\abs{\lambda_-}}^\infty dy
\frac{\sqrt{\omega_+^2-q^2} \sqrt{\frac{q^2}4 \frac{\left(q^2-\omega_+^2+4\lambda_I^2\right) \left(q^2-\omega_+^2+4\lambda_R^2\right)}{\left(q^2-\omega_+^2\right)^2} - \left[y - \frac{\omega_+}{2} \left(1+\frac{4\lambda_R\lambda_I}{q^2-\omega_+^2}\right) \right]^2}}{(y-\lambda_-)(\omega-y+\lambda_-)} \notag \\
& \hspace*{2cm} \times \heaviside{1-\left(\frac{\omega_+^2-q^2-2\omega_+y - 4\lambda_R\lambda_I}{2q\sqrt{y^2-\lambda_-^2}}\right)^2 }
\heaviside{\omega_+^2-q^2-4\gamma^2} \\
&= -\frac {g_v}8 \, \heaviside{\omega_\pm^2-q^2-4\gamma^2} \left[
\sqrt{\omega_\pm^2-q^2} 
 - \frac{\abs{q^2-\omega\left(\omega\pm2\lambda_-\right)}}{2\omega} 
 - \frac{\abs{q^2-\omega\left(\omega\pm2\lambda_+\right)}}{2\omega}
 \, \right]
\end{align}
with $\gamma = \text{max}\left\{ \lambda_R,\lambda_I \right\}$ and $y = \sqrt{k^2 + \lambda_-^2}$.

We can now make use of Eq. (\ref{KKR}) in order to find the real part. The first contribution reads
\begin{align}
&\Re{\chi^{--\to-+}(q,\omega)} = \frac 2\pi \mathcal P \int_{0}^\infty d\omega'
\frac {\omega'}{\omega'^2-\omega^2} \Im{\chi^{--\to-+} (q,\omega')} \\
&= \frac {g_v}{8\pi} \left\{ 
- K_\lambda (4\lambda^2) + \mathcal L_\lambda (\sqrt{q^2+4\lambda^2}) + 2\mathcal L_{-\lambda} (\sqrt{q^2+\lambda^2}+\lambda)
- \mathcal L_{-\lambda} (\sqrt{q^2+4\lambda^2}) \right. \notag \\
& \left. \hspace*{2cm} + \heaviside{q-\omega} \frac{3q^2\omega^2+4q^2\lambda^2-3q^4-2\omega^4}{(q^2-\omega^2)^{3/2}} \frac \pi2
\right\}
\end{align}
Here we introduced the functions
\begin{align}
K_\lambda (x) = 2\sqrt x + \frac{4q^2\lambda^2}{(q^2-\omega^2)\sqrt x} - \left(3q^4-5q^2\omega^2+2\omega^4-4q^2\lambda^2\right)
\Re{\frac{\arctan{\frac{\sqrt x}{\sqrt{q^2-\omega^2}}}}{(q^2-\omega^2)^{3/2}}}
\end{align}
and
\begin{align}
\mathcal L_\lambda (x) = x + \lambda\ln{\abs{x^2-\omega^2}} - \frac{\omega^2-q^2}{2\omega} \ln{\abs{\frac{x+\omega}{x-\omega}}}
.
\end{align}
The second contribution can be solved in a similar way,
\begin{align}
&\Re{\chi^{+-\to-+}(q,\omega)} = \frac 2\pi \mathcal P \int_{0}^\infty d\omega'
\frac {\omega'}{\omega'^2-\omega^2} \Im{\chi^{+-\to-+}(q,\omega')} \notag \\
&= -\frac{g_v}{4\pi} \left[ 2\lambda_R(1+\ln4) 
- \frac 12\Re{\sqrt{q^2-(\omega+2\lambda_R)^2} \arcsin{\frac{\omega+2\lambda_R}q}
- \sqrt{q^2-(-\omega+2\lambda_R)^2} \arcsin{\frac{-\omega+2\lambda_R}q} } \right. \notag \\
& \left. \hspace*{2cm} -\frac 12 \left[ G_{\frac\omega2 +\lambda_R}\left(\sqrt{q^2+4\gamma^2}-\omega-2\lambda_R\right)
+ G_{-\frac\omega2 +\lambda_R}\left(\sqrt{q^2+4\gamma^2}+\omega-2\lambda_R\right) \right] \right. \notag \\
& \left. \hspace*{2cm} + \heaviside{\lambda_R + \lambda_I} \mathcal L_{\lambda_-} \left(\sqrt{q^2+\lambda_-^2}-\lambda_-\right) 
- \frac 12 \sign{\lambda_R + \lambda_I} \mathcal L_{\lambda_-} \left(\sqrt{q^2+4\gamma^2}-2\lambda_R\right) \right. \notag \\
& \left. \hspace*{2cm} + \heaviside{\lambda_R - \lambda_I} \mathcal L_{\lambda_+} \left(\sqrt{q^2+\lambda_+^2}-\lambda_+\right) 
- \frac 12 \sign{\lambda_R - \lambda_I} \mathcal L_{\lambda_+} \left(\sqrt{q^2+4\gamma^2}-2\lambda_R\right)
\; \right]
\end{align}
with
\begin{align}
G_\omega (x) = & \sqrt{(x+\omega)^2-q^2} + \omega \ln{(\sqrt{(x+\omega)^2-q^2}+x+\omega)} \notag \\
& - \sqrt{\omega^2-q^2} \ln{\frac{\omega x + \omega^2-q^2 + \sqrt{\omega^2-q^2}\sqrt{(x+\omega)^2-q^2}}{x}}
\end{align}

\section{Details of the calculation of the doped polarization}
\label{sect:AppB}
The extrinsic part for the band $E_{-+}$,
\begin{align}
\delta\chi_{k_{F-}}(q,\omega) = \frac{g_v}{4\pi^2} \sum_{\mu,\nu=\pm1} \mathcal P \int_0^{k_{F-}} d^2k \sum_{\alpha=\pm1}
\frac {\alpha \left\vert \left\langle \chi_{-+}(\vec k) \Big\vert \chi_{\mu\nu}(\vec k+\vec q) \right\rangle \right\vert ^2}
{\omega + i0 - \alpha \left[ E_{\mu\nu}(\vec k+\vec q) - E_{-+}(\vec k) \right] } ,
\end{align}
can be summarized as
\begin{align}
\delta\chi_{k_{F-}}(q,\omega) =& \frac{g_v}{4\pi^2} \mathcal P \int_0^{k_{F-}} d^2k
\left[ \frac{\left(\omega + i0 + \sqrt{k^2+\lambda_+^2} + \sqrt{\abs{\vec k+\vec q}^2+\lambda_+^2}\right) 
\left\vert \left\langle \chi_{-+}(\vec k) \Big\vert \chi_{--}(\vec k+\vec q) \right\rangle \right\vert ^2
}
{\left(\omega + i0 +\sqrt{k^2+\lambda_+^2}\right)^2 - \left(\abs{\vec k+\vec q}^2 + \lambda_+^2\right)}
\right. \notag\\
& \left. + \frac{\left(\omega + i0 + \sqrt{k^2+\lambda_+^2} - \sqrt{\abs{\vec k+\vec q}^2+\lambda_+^2}\right) 
\left\vert \left\langle \chi_{-+}(\vec k) \Big\vert \chi_{--}(\vec k+\vec q) \right\rangle \right\vert ^2}
{\left(\omega + i0 + \sqrt{k^2+\lambda_+^2}\right)^2 - \left(\abs{\vec k+\vec q}^2 + \lambda_+^2\right)} \right. \notag \\
& \left. + \frac{\left(\omega_- + i0 + \sqrt{k^2+\lambda_+^2} - \sqrt{\abs{\vec k+\vec q}^2+\lambda_-^2}\right) 
\left\vert \left\langle \chi_{-+}(\vec k) \Big\vert \chi_{+-}(\vec k+\vec q) \right\rangle \right\vert ^2
}
{\left(\omega_- + i0 + \sqrt{k^2+\lambda_+^2}\right)^2 - \left(\abs{\vec k+\vec q}^2 + \lambda_-^2\right)} \right. \notag\\
& \left. + \frac {\left(\omega_- + i0 + \sqrt{k^2+\lambda_+^2} + \sqrt{\abs{\vec k+\vec q}^2+\lambda_-^2}\right) 
\left\vert \left\langle \chi_{-+}(\vec k) \Big\vert \chi_{++}(\vec k+\vec q) \right\rangle \right\vert ^2 }
{\left(\omega_- + i0 + \sqrt{k^2+\lambda_+^2}\right)^2 - \left(\abs{\vec k+\vec q}^2 + \lambda_-^2\right)} \right. \notag \\
& \left. + \left( \omega \to -\omega \right) \right. \Big] \label{doped_polarization}
\end{align}
where ($\omega \to -\omega$), and thus $(\omega_- \to -\omega_+$), denotes terms with the sign of the frequency
changed compared to the preceding expression.
The corresponding expression for $E_{++}$ can be obtained by substituting $\lambda_R \to -\lambda_R$ and $k_{F-} \to k_{F+}$.
After carrying out the angle integration for the real part and choosing a proper substitution, $x = \sqrt{k^2+\lambda_+^2} - \lambda_+ $, we arrive at
\begin{align}
&\Re{\delta\chi_{k_{F-}}(q, \omega)} = -\frac{g_v}{2\pi} \operatorname{Re} \left\{ \mathcal P \int_{\epsilon}^{\mu-\lambda_I} dx
\left[ \frac {x+\lambda_R}{2x} \right. \right. \notag \\
& \left. \left. + \frac {\left[q^2 - (x + \frac\omega2) (x+\frac\omega2 + \lambda_+) \right]^2}{x(x+ \frac \omega2)}
\frac {\sign{q^2-\omega^2-2\omega (x+\lambda_+)}} {\sqrt{q^2-\omega^2} \sqrt{\frac {q^2}4 \left(1+\frac{4\lambda_+^2}{q^2-\omega^2}\right) - (x+\frac\omega2+\lambda_+)^2}} 
\right. \right. \notag \\
& \left. \left. - \frac{\sqrt{q^2-\omega_-^2} \; \sqrt{\frac{q^2}4 \frac{\left(q^2-\omega_+^2+4\lambda_I^2\right) \left(q^2-\omega_+^2+4\lambda_R^2\right)}{\left(q^2-\omega_+^2\right)^2} - (x + \frac{\omega_+}{2} \left(1+\frac{4\lambda_R\lambda_I}{q^2-\omega_+^2}\right) + \lambda_+)^2}} {4x(x+\omega)} \times \right. \right. \notag \\
& \left. \left. \hspace*{1cm} \times \sign{q^2-\omega_-^2 -2\omega_- (x+\lambda+) - 4\lambda_R \lambda_I } \right] 
\quad + \quad \left( \omega \to -\omega \right) \; \right. \Big\}
\end{align}
These integrals can now be solved in terms of trigonometric and hyperbolic functions.\cite{Gradshteyn}
In order to simplify the expressions we use the shorthand notation \cite{Gamayun_2011}
\begin{align}
\hat f(x)\Big\vert_a^b = \sign{b-x} \left( f(b) - f(x) \right) - \sign{a-x} \left( f(a) - f(x) \right)
\end{align}
The result can then be written as
\begin{align}
& \Re{\delta\chi_{k_{F-}}(q,\omega)}  = - \frac{g_v\left(\mu-\lambda_I\right)}{2\pi} - \frac{g_v\lambda_R}{4\pi}\ln{\frac{\mu-\lambda_I}\epsilon} \notag \\
& + \frac{g_v}{2\pi\omega} \operatorname{Re} \left\{
\sign{\omega} \left(
\hat{\mathcal R}_1^\omega \left(\frac{q^2 - \omega^2 - 2\omega\lambda_+}{2\omega}\right) \Big\vert_{\epsilon}^{\mu-\lambda_I} 
- \hat{\mathcal R}_1^{-\omega} \left(\frac{q^2 + \omega^2 - 2\omega\lambda_+}{2\omega}\right) \Big\vert_{\omega}^{\mu-\lambda_I+\omega} 
\right) \right. \notag \\
& \left. - \sign{\omega_-} 
\left( \hat{\mathcal R}_2^\omega \left(\frac{q^2-\omega^2+2\lambda_-\omega}{2\omega_-}\right) \Big\vert_{\epsilon}^{\mu-\lambda_I}
- \hat{\mathcal R}_2^{-\omega} \left(\frac{q^2+\omega^2-2\lambda_+\omega}{2\omega_-}\right) \Big\vert_{\omega}^{\mu-\lambda_I+\omega}
\right) \right\}
\quad + \quad \left( \omega \to - \omega \right)
\label{extrinsic_part_final}
\end{align}
with
\begin{align}
\mathcal R_i^{\omega}(x) = \frac {c_i^\omega}{\gamma_i^\omega} \sqrt{r_i^\omega} 
- \frac{\sqrt{\alpha_i^\omega}}4 \ln{\frac{2\alpha_i^\omega + \beta_i^\omega x + 2\sqrt{\alpha_i^\omega} \sqrt{r_i^\omega}}x}
+ \frac {\tilde {c}_i^\omega}{\sqrt{\gamma_i^\omega}} \ln{\left(2\sqrt{\gamma_i^\omega} \sqrt{r_i^\omega} + 2\gamma_i^\omega x + \beta_i^\omega \right)}
\end{align}
and $r_i^\omega = \alpha_i^\omega + \beta_i^\omega x + \gamma_i^\omega x^2$. The coefficients read
\begin{align*}
\alpha_1^\omega &= \frac{\left(q^2 - \omega\left(\omega+2\lambda_+\right)\right)^2}{4} \quad , \quad
\alpha_2^\omega = \frac{\left(q^2 - \omega\left(\omega_- - 2\sign{\omega}\lambda_I\right)\right)^2}{4}
\\
&\beta_1^\omega = \left(\omega^2-q^2\right) \left( 2\lambda_+ + \omega \right) \quad , \quad
\beta_2^\omega = 8\lambda_R^2\lambda_I - 2\lambda_R\lambda_I\left(\omega+\abs{\omega}\right) - \sign{\omega}\left(\omega-2\lambda_R\right) \left(q^2-\omega_-^2\right)
\\
\gamma_1^\omega &= \omega^2-q^2 \quad , \quad
\gamma_2^\omega = \abs{\omega}_-^2 - q^2 \quad , \quad
c_1^\omega = \frac {x^2}3 + \frac{2\lambda_+ + 7\omega}{12} x + \frac{4\lambda_+^2 - 4q^2 + 8\lambda_+^2 + 5\omega^2}{8} - \frac {2\alpha_1^\omega}{3\gamma_1^\omega} \\
\tilde c_1^\omega &= \frac{\alpha_1^\omega \left(2\lambda_+ - \omega\right)}{4\gamma_1^\omega}
+ \frac {3\omega^3 + 6\lambda_+^2\omega^2 - 4(q^2+\lambda_+^2)\omega - 8\lambda_+^3}{16} \quad , \quad
c_2^\omega = \frac{\gamma_2^\omega}4 \quad , \quad \tilde c_2^\omega = \frac{\beta_2^\omega}{8}
\end{align*}

The calculation of the imaginary part is quite similar. 
Starting from Eq. (\ref{doped_polarization}) and carrying out the angle integration in a way similar to the real part, we arrive at
\begin{align}
&\Im{\delta\chi_{k_{F-}}(q, \omega)} = \frac{g_v}{2\pi} \operatorname{Re} \left\{ \mathcal P \int_{\epsilon}^{\mu-\lambda_I} dx
\left[ \frac {\left[q^2 - (x + \frac\omega2) (x+\frac\omega2 + \lambda_+) \right]^2}{x(x+ \frac \omega2)} 
\frac {\sign{x-\lambda_-+\omega}} {\sqrt{\omega^2-q^2} \sqrt{\frac {q^2}4 \left(1+\frac{4\lambda_+^2}{q^2-\omega^2}\right) - (x+\frac\omega2+\lambda_+)^2}} \right. \right. \notag \\
& \left. \left. - \frac{\sqrt{\omega_-^2-q^2} \sqrt{\frac{q^2}4 \frac{\left(q^2-\omega_+^2+4\lambda_I^2\right) \left(q^2-\omega_+^2+4\lambda_R^2\right)}{\left(q^2-\omega_+^2\right)^2} - (x + \frac{\omega_+}{2} \left(1+\frac{4\lambda_R\lambda_I}{q^2-\omega_+^2}\right) + \lambda_+)^2}} {4x(x+\omega)} 
 \sign{x+\lambda_+ + \omega} \; \right] 
\quad - \quad \left( \omega \to -\omega \right) \; \right. \Big\}
\end{align}
The result can again be written as
\begin{align}
&\Im{\delta\chi_{k_{F-}} (q,\omega)} = 
\frac{g_v}{2\pi\omega} \operatorname{Re} \left\{
\frac 1i \hat{\mathcal R}_1^\omega \left(-\lambda_+ - \omega\right) \Big\vert_{\epsilon}^{\mu-\lambda_I}
- \frac 1i \hat{\mathcal R}_1^\omega \left(-\lambda_+\right) \Big\vert_{\omega}^{\mu-\lambda_I+\omega}
\right. \notag \\
&\left. -\frac 1i \heaviside{-\omega}\heaviside{\mu-\lambda_I+\omega}\sign{\lambda_+} 
\left[ R_1^\omega (-\epsilon) - R_1^\omega (\epsilon) \right]
+ i \hat{\mathcal R}_2^\omega \left(\lambda_--\omega\right) \Big\vert_{\epsilon}^{\mu-\lambda_I}
- i \hat{\mathcal R}_2^\omega \left(\lambda_-\right) \Big\vert_{\omega}^{\mu-\lambda_I+\omega}
\right. \notag \\
& \left.
-i  \heaviside{-\omega}\heaviside{\mu-\lambda_I+\omega}\sign{-\lambda_-}
\left[ R_2^\omega (-\epsilon) - R_2^\omega (\epsilon) \right]
\right\}
\quad - \quad \left( \omega \to - \omega \right) \label{extrinsic_im_part_final}
\end{align}
The limit $\epsilon\to0$ in Eqs. (\ref{extrinsic_part_final}) and (\ref{extrinsic_im_part_final}) can now be taken safely giving finite results.

\end{widetext}


\end{document}